\documentclass[floatfix,pra,twocolumn,showpacs,amsmath,amssymb,letterpaper,groupaddresses,superscriptaddress]{revtex4}
\usepackage{times}
\usepackage{latexsym}
\usepackage{graphicx}
\usepackage{verbatim,times,bbm}
\usepackage{color}

\usepackage{color}

\def\be{\begin{equation}}
\def\ee{\end{equation}}
\def\ba{\begin{eqnarray}}
\def\ea{\end{eqnarray}}
\def\la{\langle}
\def\ra{\rangle}

\begin{document}

\title{Entanglement dynamics for qubits dissipating into a common environment}

\author{Laleh Memarzadeh\footnote{email: memarzadeh@sharif.edu}}
\affiliation{Department of Physics, Sharif University of Technology, Teheran, Iran}

\author{Stefano Mancini\footnote{email: stefano.mancini@unicam.it}}
\affiliation{School of Science and Technology, University of Camerino, I-62032 Camerino, Italy}
\affiliation{INFN-Sezione di Perugia, I-06123 Perugia, Italy}

\begin{abstract}
We provide an analytical investigation of the entanglement dynamics for a system composed of an arbitrary number
of qubits dissipating into a common environment. Specifically we consider initial states whose evolution remains confined on low dimensional subspaces of the operators space. We then find for which pairs of qubits entanglement can be generated and can persist at steady state. Finally, we determine the stationary distribution of entanglement as well as its scaling versus the total number of qubits in the system.

\end{abstract}

\pacs{03.67.Bg, 03.65.Yz}

\maketitle

\section{Introduction}

Entanglement is synonymous of quantum correlations that cannot be explained by any local classical theory.
Initially this notion was relegated to foundational issues, but in the last decades  it has pervaded quantum information and other areas of physics \cite{entrev}. Being a purely quantum phenomenon it was considered fragile under contamination of environment noise in open quantum systems. Nevertheless, in recent years it has been shown that it can persist for long time up to stationary conditions in several contexts \cite{clark03} (see also \cite{polzik10} for a striking experiment on long living entanglement). More importantly it has been shown that the environment can play a constructive role in establishing entanglement \cite{eassist,braun02}.  Quite remarkably it happens that even without any interaction among subsystems a common dissipative environment is able to induce entanglement \cite{braun02}. However, this possibility has been proved true only for systems composed by two subsystems due to technical difficulties arising when accounting for more subsystems.

Here we overcome these difficulties and achieve results for an arbitrary number of subsystems by considering a limited number of initially available excitations and by exploiting a power expansion for the dissipation superoperator. This allows us to restrict the analysis to a low dimensional subspace of the operators space.

We find that pairwise entanglement is not created at any time between initially excited qubits.
Instead, it is created and persists at steady state for pairs of initially excited and initially not excited qubits.
The same holds true for pairs of qubits initially in the ground state, but in such a case the amount of entanglement is negligibly smaller than the previous case.

%%%%%%%%%%%%%%%%%%%%%%%%%%%%%%%%%%%%%%%%%%%%%%%%%%%%%%%%%%%%

\section{System Dynamics}

Let us consider a system of $n$ qubits with associated Hilbert space $\mathcal{H}\simeq\mathbb{C}^{2\otimes n}$. Let $\{|0\ra,|1\ra\}^{\otimes n}$ be the orthonormal basis, with $|0\rangle$ (resp. $|1\rangle$) the ground (resp. excited) single qubit state.

We want to study the dynamics of this system of qubits when they dissipate together into an environment
at zero temperature.
It will be governed by the Lindblad master equation \cite{qnoise}:
\begin{equation}\label{me}
\dot\rho(t)=2\sigma \rho(t) \sigma^{\dag} -\sigma^\dag \sigma\rho(t)- \rho(t)\sigma^\dag \sigma
\equiv\mathcal{D}\rho(t),
\end{equation}
where
\be
\sigma:=\sum_{i=1}^n\sigma_i,
\ee
with $\sigma_i:=|0\ra\la 1|$ for the $i$th qubit.
The dissipation rate has been set equal to $1$ for the sake of simplicity.

Let us describe the strategy we will put into practice to solve \eqref{me}.
The fomal solution reads $\rho(t)=e^{t\mathcal{D}}\rho(0)$.
To explicitly find  $\rho(t)$ let us first write the Taylor expansion:
\begin{equation}\label{Taylor1}
\rho(t)=\rho(0)+t\mathcal{D}\rho(0)+\frac{t^2}{2!}\mathcal{D}^2\rho(0)+\frac{t^3}{3!}\mathcal{D}^3\rho(0)+\cdots .
\end{equation}
Now notice that once the initial state $\rho(0)$ has been chosen, the dynamics (being purely dissipative at zero temperature) can only decrease the number of initial excitation present in $\rho(0)$.
That is, repeated applications of $\mathcal{D}$ to $\rho(0)$ will leave the state within a
subspace $\mathbb{H}_{\rho(0)}\subset \mathbb{H}$ of the Hilbert space  $\mathbb{H}=\mathcal{H}\otimes\mathcal{H}^*$ where $\mathcal{H}^*$ stands for the dual of $\mathcal{H}$.
After having identified $\mathbb{H}_{\rho(0)}$, i.e. a set of operators on $\mathcal{H}$ spanning  $\mathbb{H}_{\rho(0)}$, we will write down $\rho(0)$ as linear combination of such operators with unknown time dependent coefficients.
Then, by inserting this expansion into Eq.\eqref{me} we will derive a set of linear differential equations for the unknown coefficients. In this way if the initial state contains a small number of excitation we can hope to provide an analytical solution. In fact given a number of initial excitations $e$ ($e\ll n$) the following inequality holds true:
\be
dim \mathbb{H}_{\rho(0)} \le \left( 2^e \right)^2 \ll \left( 2^{n} \right)^2 = dim \mathbb{H}.
\label{eqdim}
\ee

Finally, notice that \eqref{me} does not satisfy the condition for the uniqueness of a stationary solution.
This condition requires that the only operators commuting with the
Lindblad operator $\sigma$ must be multiples of identity \cite{Spohn}. Hence we should expect different steady states depending on the choice of $\rho(0)$.

%%%%%%%%%%%%%%%%%%%%%%%%%%%%%%%%%%%%%%%%%%%%%%%%%%%%%%%%%%%%

\section{One initial excitation ($e=1$)}

In this section we assume that initially only the $k$th qubit is in the excited state while all the others are in the ground state. Let us denote this state as $|k\rangle$, then $\rho(0)=|k\ra\la k|$.

By applying $\mathcal{D}$ to $\rho(0)=|k\ra\la k|$
we can easily find the following closed relations:
\begin{eqnarray}
\mathcal{D}|k\ra\la k|&=&2|G\ra\la G|-(|E_{\not k}\ra\la k|+|k\ra\la E_{\not k}|),\notag\\
\mathcal{D}|G\ra\la G|&=&0,\notag\\
\mathcal{D}\left(|E_{\not k}\ra\la k|+|k\ra\la E_{\not k}|\right)&=&4(n-1)|G\ra\la G|-2|E_{\not k}\ra\la E_{\not k}|\notag\\
&&-2(n-1)|k\ra\la k|\notag\\
&&-n(|E_{\not k}\ra\la k|+|k\ra\la E_{\not k}|),\notag\\
\mathcal{D}|E_{\not k}\ra\la E_{\not k}|&=&2(n-1)^2|G\ra\la G|\notag\\
&&-2(n-1)|E_{\not k}\ra\la E_{\not k}|\notag\\
&&-(n-1)(|E_{\not k}\ra\la k|+|k\ra\la E_{\not k}|),\notag\\
\label{cr1}
\end{eqnarray}
where we have defined
\begin{eqnarray}
\label{defGE}
|G\ra:=|0\ra^{\otimes n},\qquad
|E_{\not k}\ra:=\sum_{i\neq k}^n|i\ra,
\end{eqnarray}
($|i\ra$ stands for a state in which the $i$th qubit is in the excited state and all the others are in the ground state).

With the help of Eq.\eqref{cr1} we identify the subspace
\ba
\mathbb{H}_{\rho(0)}&=&span\left\{ |G\ra\la G|, |k\ra\la k|, |E_{\not k}\ra\la E_{\not k}|, \right.\notag\\
&&\left. \hspace{1cm} (|E_{\not k}\ra\la k|+|k\ra\la E_{\not k}|) \right\}.
\ea
Then we can write
\ba
\rho(t)&=&a_0(t)  |G\ra\la G| +a_1(t) |k\ra\la k| +a_2(t) |E_{\not k}\ra\la E_{\not k}| \notag\\
&&+a_3(t) (|E_{\not k}\ra\la k|+|k\ra\la E_{\not k}|),
\ea
which, upon insertion into \eqref{me}, leads to the set of differential equations
\ba
\dot{a}_0&=&2a_1+2(n-1)^2a_2+4(n-1)a_3 ,\notag\\
\dot{a}_1&=&-2a_1-2(n-1)a_3 ,\notag\\
\dot{a}_2&=&-2(n-1)a_2-2a_3 ,\notag\\
\dot{a}_3&=&-a_1-(n-1)a_2-n a_3 ,
\ea
with initial conditions
$a_0(0)=a_2(0)=a_3(0)=0, a_1(0)=1$.

Solving them we arrive at the density operator
\begin{eqnarray}\label{rt}
\rho(t)&=&(1-f(t))^2|k\ra\la k|+f(t)(2-nf(t))|G\ra\la G|\notag\\
&&+f(t)^2|E_{\not k}\ra\la E_{\not k}|\notag\\
&&-f(t)(1-f(t))(|E_{\not k}\ra\la k|+|k\ra\la E_{\not k}|),
\end{eqnarray}
where $f(t):=\frac{1}{n}(1-e^{-nt})$.

To analyze the pairwise entanglement between the qubit $k$ and a generic other qubit $j$
we compute the partial trace of \eqref{rt} overall the other qubits and obtain
the following density operator
\begin{eqnarray}
\rho_{k,j}&=&(1-f(t))^2|10\ra\la 10|+f^2(t)|01\ra\la 01|\notag\\
&&-f(t)(1-f(t))\left(|10\ra\la 01|+|01\ra\la 10|-2|00\ra\la 00|\right).\notag\\
\label{rhokj1}
\end{eqnarray}

\begin{figure}[t]
  \centering
  \includegraphics[scale=0.35]{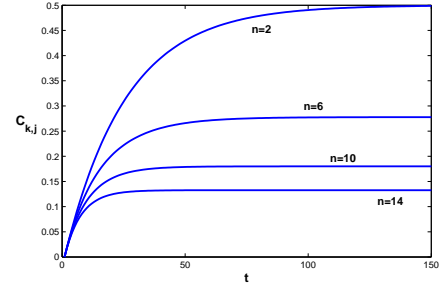}
  \caption{Concurrence $C_{k,j}$ versus time $t$. From top to bottom the system size is $n=2,6,10,14$.}
    \label{C1Excitation}
\end{figure}

Using the concurrence \cite{Wootters} as a measure of entanglement of  $\rho_{k,j}$ in Eq.(\ref{rhokj1})
we find its time evolution as
\begin{equation}\label{CkjT}
C_{k,j}(t)=\frac{2}{n}\left(1-e^{-nt}\right)\left(1-\frac{1-e^{-nt}}{n}\right).
\end{equation}
As it is shown in Fig.\ref{C1Excitation}, entanglement is generated between the $k$th qubit (initially in the excited state) and an arbitrary qubit $j$ (initially in the ground state), although there is no direct interaction between them.
The maximum amount of entanglement decreases with the system size $n$, but the concurrence achieves its maximum value faster when the system size is larger.

%%%%%%%%%%%%%%%%%%%%%%%%%%%%%%%%%%%%%%%%%%
%%%%%%%%%%%%%%%%%%%%%%%%%%%%%%%%%%%%%%%%%%

\begin{figure}[t]
  \centering
    \includegraphics[scale=0.35]{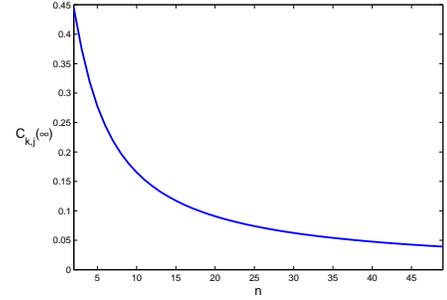}
  \caption{Stationary concurrence $C_{k,j}$ versus system size $n$.}
    \label{C1ExcitationStationary}
\end{figure}

Letting $t$ going to infinity in Eq.(\ref{CkjT}) we can find the behavior of stationary entanglement versus
the system size $n$ as
\begin{equation}
C_{k,j}(\infty)=\frac{2(n-1)}{n^2}.
\end{equation}
This is shown in Fig.\ref{C1ExcitationStationary}.

%%%%%%%%%%%%%%%%%%%%%%%%%%%%%%%%%%%%%%%%%%
%%%%%%%%%%%%%%%%%%%%%%%%%%%%%%%%%%%%%%%%%%

To study the entanglement between any two qubits initially in the ground state, we compute the reduced density matrix of $j$th and $m$th qubits ($j,m\neq k$). By referring to Eq.(\ref{rt}) it is easy to show that
\ba
\rho_{j,m}&=&\left(1-2f^2(t)\right)|00\ra\la 00|\notag\\
&&+f^2(t)\left(|10\ra+|01\ra\right)(\left(\la 10|+\la 01|\right).
\label{rhojm1}
\ea

\begin{figure}[t]
  \centering
  \includegraphics[scale=0.35]{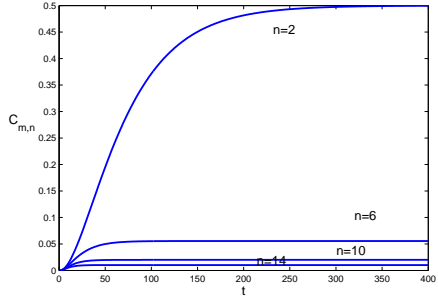}
  \caption{Concurrence $C_{j,m}$ versus time $t$. From top to bottom the system size is $n=2,6,10,14$.}
    \label{ConcurrenceMN1excitation}
\end{figure}

\begin{figure}[t]
  \centering
   \includegraphics[scale=0.38]{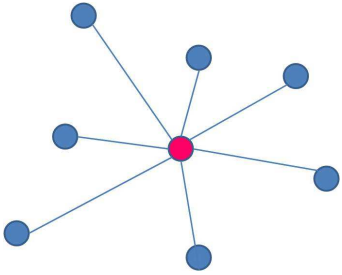}
  \caption{Pictorial representation of the leading stationary correlations (edges) among qubits (circles).
   Red (reps. blue) circles represents qubits initially in the excited (resp. ground) state.}
    \label{Star}
\end{figure}

The concurrence \cite{Wootters} of $\rho_{j,m}$ in Eq.(\ref{rhojm1}) results
\begin{equation}\label{CmnT}
C_{_{j,m}}(t)=2\left(\frac{1-e^{-nt}}{n}\right)^2.
\end{equation}
Its behavior is shown in Fig.\ref{ConcurrenceMN1excitation}. From Eq.(\ref{CmnT}) it is easy to see that at steady state it is $C_{j,m}(\infty)=\frac{2}{n^2}$, which is negligible compared to $C_{k,j}(\infty)$
for large $n$. Therefore for large systems, we have a star graph (see Fig.\ref{Star}) as steady state.

%%%%%%%%%%%%%%%%%%%%%%%%%%%%%%%%%%%%%%%%%%
%%%%%%%%%%%%%%%%%%%%%%%%%%%%%%%%%%%%%%%%%%

%%%%%%%%%%%%%%%%%%%%%%%%%%%%%%%%%%%%%%%%%%
%%%%%%%%%%%%%%%%%%%%%%%%%%%%%%%%%%%%%%%%%%%%%%

%%%%%%%%%%%%%%%%%%%%%%%%%%%%%%%%%%%%%%%%%%%%%%%%%%%%%%%%%%%%%%%%%%%%%%

\section{Two initial excitations ($e=2$)}

In this section we assume that initially two qubits, say the $k$th and the $l$th, are in the excited state,
while all the others are in the ground state. Let us denote this state as $|k,l\rangle$, then $\rho(0)=|k,l\ra\la k,l|$.

By applying $\mathcal{D}$ to $\rho(0)=|k,l\ra\la k,l|$
we can find the following closed relations:
\begin{widetext}
\begin{eqnarray}\label{l}
\mathcal{D}|k,l\ra\la k,l|&=&2|k+l\ra\la k+l|-4|k,l\ra\la k,l|-\Lambda,\notag\\
\mathcal{D}|k+l\ra\la k+l|&=&8|G\ra\la G|-2\Omega-4|k+l\ra\la k+l|,\notag\\
\mathcal{D}|G\ra\la G|&=&0,\notag\\
\mathcal{D}\Omega&=&2(n-2)\left( 4|G\ra\la G|-|k+l\ra\la k+l| \right) -n\Omega-4|E_{\not k\not l}\ra\la E_{\not k\not l}|,\notag\\
\mathcal{D}|E_{\not k\not l}\ra\la E_{\not k\not l}|&=&(n-2)\left(2(n-2)|G\ra\la G|-\Omega-2| E_{\not k\not l}\ra\la E_{\not k\not l}|\right),\notag\\
\mathcal{D}\Lambda&=&4(n-2)\left(|k+l\ra\la k+l|-|k,l\ra\la k,l|\right)-2|B\ra\la B|+4(\Omega-\Pi)-(n+2)\Lambda,\notag\\
\mathcal{D}| B\ra\la B|&=&2(n-2)^2|k+l\ra\la k+l|+2(n-2)(2\Omega-\Lambda)+8| E_{\not k\not l}\ra\la E_{\not k\not l} |-2n |B\ra\la B|-4\Gamma,\notag\\
\mathcal{D}\Pi&=&(n-3)\left(2\Omega-\Lambda\right)-2(n-2)\Pi-\Gamma,\notag\\
\mathcal{D}\Gamma&=&2(n-3)\left(4| E_{\not k\not l} \ra\la E_{\not k\not l} |-|B\ra\la B|\right)+2(n-2)(n-3)\Omega-8|H\ra\la H|-2(n-2)\Pi-3(n-2)\Gamma,\notag\\
\mathcal{D}|H\ra\la H|&=&(n-3)\left(2(n-3) | E_{\not k\not l}\ra\la E_{\not k\not l} |-4 |H\ra\la H|-\Gamma\right),
\label{cr2}
\end{eqnarray}
\end{widetext}
where, in addition to Eq.\eqref{defGE}, we have defined
\begin{eqnarray}
|k+l\ra&:=&|k\ra+|l\ra, \notag\\
| E_{\not k\not l} \ra&:=&\sum_{i\neq k,l}^n|i\ra, \notag\\
|B\ra&:=&\sum_{i\neq k}|i,k\ra+\sum_{i\neq l}|i,l\ra, \notag\\
|H\ra&:=&\sum_{h>i}|h,i\ra, \hskip 2cm h,i\neq k,l, \notag\\
\end{eqnarray}
and
\begin{eqnarray}
\Omega&:=&| E_{\not k\not l}\ra\la k+l|+|k+l\ra\la E_{\not k\not l}|, \notag\\
\Lambda&:=&|B\ra\la k,l|+|k,l\ra\la B|, \notag\\
\Pi&:=&|H\ra\la k,l|+|k,l\ra\la H|, \notag\\
\Gamma&:=&|B\ra\la H|+|H\ra\la B|.
\end{eqnarray}
With the help of Eq.\eqref{cr2} we can identify the subspace
\ba
\mathbb{H}_{\rho(0)}&=&span\left\{ |G\ra\la G|, |E_{\not k\not l}\ra\la E_{\not k\not l}|, |k+l\ra\la k+l|,\Omega, \right.\notag\\
&&\left. \hspace{1cm} |k,l\ra\la k,l|, |B \ra\la B|, |H\ra\la H|, \Lambda,\Pi, \Gamma \right\}.\notag\\
\ea
Then we can write
\begin{eqnarray}
\rho(t)&=&b_0(t) |G\ra\la G|+b_1(t) | E_{\not k\not l}\ra\la E_{\not k\not l} |+b_2(t) |k+l\ra\la k+l|\notag\\
&&+b_3(t) \Omega+b_4(t) |k,l\ra\la k,l| +b_5(t) |B\ra\la B|\notag\\
&&+b_6(t) |H\ra\la H|+b_7(t) \Lambda+b_8(t) \Pi+b_9(t) \Gamma.
\label{rhoexpb}
\end{eqnarray}
By inserting this expansion of $\rho(t)$ into Eq.\eqref{me} we arrive at the following set of differential equations:
\ba
\dot{b}_0(t)&=&2(n-2)^2b_1(t)+8b_2(t)+8(n-2)b_3(t),\label{eqb0}\\
\dot{\bf v}(t)&=&M {\bf v}(t),
\label{eqv}
\ea
where ${\bf{v}}=(b_1,b_2,b_3,b_4,b_5,b_6,b_7,b_8,b_9)^{\sf T}$ and
\begin{widetext}
\begin{equation}
M=\left(\begin{array}{cccccccccc}
   -2(n-2)&0&-4&0&8&2(n-3)^2&0&0&8(n-3)\cr
   0&-4&-2(n-2)&2&2(n-2)^2&0&4(n-2)&0&0\cr
   -(n-2)&-2&-n&0&4(n-2)&0&4&2(n-3)&2(n-2)(n-3)\cr
   0&0&0&-4&0&0&-4(n-2)&0&0\cr
   0&0&0&0&-2n&0&-2&0&-2(n-3)\cr
   0&0&0&0&0&-4(n-3)&0&0&-8\cr
   0&0&0&-1&-2(n-2)&0&-(n+2)&-(n-3)&0\cr
   0&0&0&0&0&0&-4&-2(n-2)&-2(n-2)\cr
   0&0&0&0&-4&-(n-3)&0&-1&-3(n-2)
   \end{array}\right),
\end{equation}
\end{widetext}
with the initial conditions $b_0(0)=0$, ${\bf v}(0)=(0,0,0,1,0,0,0,0,0)^{\sf T}$.

It is worth remarking that Eq.\eqref{eqb0} has been kept out of Eq.\eqref{eqv}
to make the matrix $M$ non-singular.
Still $M$ is not normal, hence cannot be diagonalized.
Nevertheless Eq.\eqref{eqv} can be solved using the Laplace transform
\ba
\tilde{\bf v}(s)&:=&\int_0^{\infty}e^{-st}{\bf v}(t)dt\notag\\
&=&(s I_{9\times 9}-M)^{-1}{\bf v}(0),
\ea
where $I_{9\times 9}$ denotes the $9\times 9$ identity matrix.
By inverting the Laplace transform we get  the coefficients
$b_i,\;i=1,\ldots 9$. Then, $b_0$ straightforwardly follows by integrating Eq.\eqref{eqb0}.
Their explicit expressions are given in Appendix A.

The reduced density operator of qubits $k$ and $l$ (initially in excited states) results from
Eq.\eqref{rhoexpb} as
\begin{eqnarray}
&&\rho_{_{k,l}}(t)=b_{_4}(t)|11\ra\la 11|\notag\\
&&+\left(b_{_2}(t)+(n-2)b_{_5}(t)\right)(|10\ra+|01\ra)(\la 10|+\la 01|)\notag\\
&&+\left(b_{_0}(t)+(n-2)\left(b_{_1}(t)+\frac{(n-3)}{2}b_{_6}(t)\right)\right)|00\ra\la 00|.\notag\\
\end{eqnarray}
By computing the concurrence \cite{Wootters} of $\rho_{k,l}(t)$ it is possible to see that the two qubit are not entangled at any time.

\begin{figure}]t]
  \centering
  \includegraphics[scale=0.35]{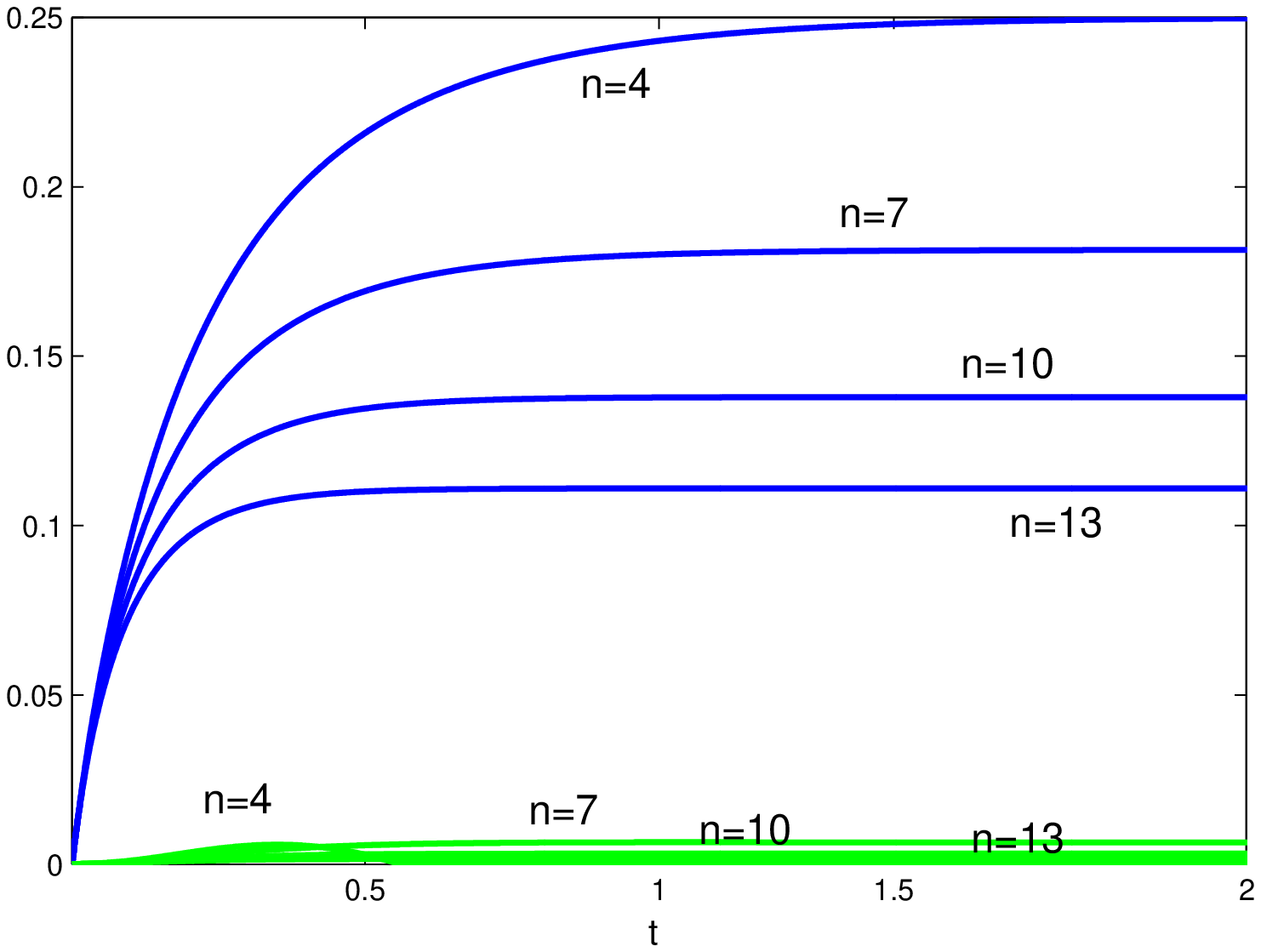}
  \caption{Blue lines: concurrence $C_{k,j}$ versus time $t$. Green Lines: concurrence $C_{j,m}$ versus time $t$. From top to bottom the system size is $n=4,7,10,13$.}
    \label{CkmCmn}
\end{figure}

The reduced density operator of qubit $k$ and a generic qubit $j$ initially in the ground state
derived from Eq.\eqref{rhoexpb} is given by
\begin{eqnarray}
\rho_{_{k,j}}(t)&=&b
_{_5}(t)|11\ra\la 11|\notag\\
&+&\left(b_{_2}(t)+(n-3)b_{_5}(t)+b_{_4}(t)\right)|10\ra\la 10|\notag\\
&+&(b_{_1}(t)+b_{_5}(t)+(n-3)b_{_6}(t))|01\ra\la 01|\notag\\
&+&(b_{_3}(t)+(n-3)b_{_9}(t)(|10\ra\la 01|+|01\ra\la 10|)\notag\\
&+&\Big(b_{_0}(t)+(n-3)b_{_1}(t)+b_{_2}(t) +(n-3)b_{_5}(t) \notag\\
&&+\frac{(n-3)(n-4)}{2}b_{_6}(t)\Big) |00\ra\la 00|.
\label{rhokl2}
\end{eqnarray}
Then, the concurrence \cite{Wootters} of the state (\ref{rhokl2}) results
\begin{eqnarray}
C_{k,j}(t)&=&-2(b_{3}(t)+(n-3)b_{9}(t)+b_{7}(t))\notag\\
&&-2\Big[b_{5}(t)(b_{0}(t)+(n-3)b_{1}(t)+b_{2}(t)\notag\\
&&+(n-3)b_{5}(t)+\frac{(n-3)(n-4)}{2}b_{6}(t)\Big]^{1/2}.\notag\\
\label{Ckj}
\end{eqnarray}
Blue lines in Fig.\ref{CkmCmn} show
the behavior of entanglement between qubits $k$ and $j$ quantified by $C_{k,j}$ vs time.
It is generated as qubits start interacting with the environment and,
as it can be seen, the stationary entanglement is achieved faster as the system size increases, 
though the maximum value decreases.
In the limit of $t\to \infty$ Eq.(\ref{Ckj}) becomes
\ba
C_{k,j}(\infty)&=&\frac{2(n^5-11n^4+39n^3-53n^2+24n-8)}{n^2(n-1)^2(n-2)^2}\notag\\
&-&\frac{2(n-3)}{n(n-1)^2(n-2)^2}\Big[7n^4-50n^3+119n^2\notag\\
&&-104n+22\Big]^{1/2} ,\notag\\
\ea
which for large $n$ behaves like $\frac{2}{n}$.

%%%%%%%%%%%%%%%%%%%%%%%%%%%%%%%%%%%%%%%%%%%%%%%%%%%%%%%%%%%%%%
%%%%%%%%%%%%%%%%%%%%%%%%%%%%%%%%%%%%%%%%%%%%%%%%%%%%%%%%%%%%%%%
The other possible case study is the entanglement between qubits initially in ground sate, 
say qubits $j$ and $m$ ($j,m\neq k,l$). Their reduced density operator, 
derived from Eq.\eqref{rhoexpb}, reads
\begin{eqnarray}
&&\rho_{_{j,m}}=b_{_6}(t)|11\ra\la 11|\notag\\
&&+(b_{_1}(t)+2b_{_5}(t)+(n-4)b_{_6}(t))\notag\\
&&\times(|10\ra+|01\ra) (\la 10|+\la 01|)\notag\\
&&+\Big(b_{_0}(t)+(n-4)b_{_1}(t)+2b_{_2}(t)+b_{_4}(t) \notag\\
&&+2(n-4)b_{_5}(t)+\frac{(n-4)(n-5)}{2}b_{_6}(t)\Big)|00\ra\la 00|.\notag\\
\label{rhojm2}
\end{eqnarray}

\begin{figure}[t]
  \centering
  \includegraphics[scale=0.25]{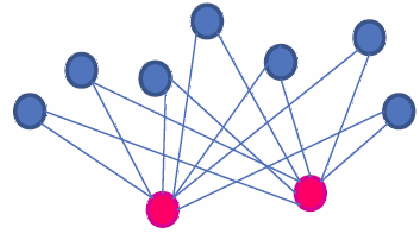}
  \caption{Pictorial representation of the leading stationary correlations (edges) among qubits (circles). Red (reps. blue) circles represents qubits initially in the excited (resp. ground) state.}
    \label{TwoColorGraph}
\end{figure}

The concurrence \cite{Wootters} of the state (\ref{rhojm2}) reads
\ba
C_{j,m}(t)&=&2(b_{1}(t)+2b_{5}(t)+(n-4)b_{6}(t))\notag\\
&-&2\Big[b_{6}(t)(b_{0}(t)+(n-4)b_{1}(t)+2b_{2}(t)+b_{4}(t) \notag\\
&&+2(n-4)b_{5}(t)+\frac{(n-4)(n-5)}{2}b_{6}(t))\Big]^{1/2}.\notag\\
\ea
In the stationary condition $C_{j,m}$ becomes
\ba
C_{j,m}(\infty)&=&\frac{4(n^4-2n^3-7n^2+10n-4)}{n^2(n-1)^2(n-2)^2}\notag\\
&-&\frac{4}{n(n-1)^2(n-2)^2}\Big[n^6-10n^5+41n^4\notag\\
&&-104n^3+180n^2-152n+48\Big]^{1/2} ,
\ea
which vanishes for large system size.
Green lines in Fig.\ref{CkmCmn} show that the amount of entanglement generated between qubits initially in the ground state is negligible compared to that of qubits $k$ and $j$. Therefore
in the stationary state we have a bipartite graph as depicted in Fig.\ref{TwoColorGraph}.
%%%%%%%%%%%%%%%%%%%%%%%%%%%%%%%%%%%%%%%%%%
%%%%%%%%%%%%%%%%%%%%%%%%%%%%%%%%%%%%%%%%%%%%%%

\section{Concluding Remarks}

In conclusion, we have studied the entanglement dynamics of a system composed of an arbitrary number
of qubits dissipating into a common environment. 

The consideration of models like this is becoming increasingly
pressing with the continuing miniaturization of
information processing devices.
For instance, in solid state implementations qubits may be so closely spaced that
they can experience the same environment \cite{Hu}.

By considering initial states whose evolution is confined on low dimensional subspaces of the operators space
we found that pairwise entanglement is not created at any time between initially excited qubits.
Instead, it is created and persists at steady state for pairs of initially excited and initially not excited qubits.
The same holds true for pairs of qubits initially in the ground state, but in such a case the amount of entanglement is negligibly smaller than the previous case.
This lead to the emergence of structures (subgraphs) which makes this subject appealing also from the perspective of  complex networks \cite{Barab}.

Finally, it is worth noticing that the extension of the presented analysis 
to states with an arbitrary number $e$ of initial excitations
can be done in a computationally efficient way due to the polynomial scaling of 
$dim\mathbb{H}_{\rho(0)}$ vs $e$.
We can arrive at such a scaling (much tighter
with respect to the bound given in Eq.(\ref{eqdim})),
by simply observing the results in Sec.III and IV. 
In fact, in Sec.III  the density matrix can be written, in the vector basis
$\{|G\rangle, |k\rangle, |E_{\not k}\rangle\}$, as
\begin{eqnarray}
\rho=\left(\begin{array}{ccc}
a_0 & 0 & 0 \\
0 & a_1 & a_3\\
0 & a_3 & a_2
\end{array}\right).
\label{rhoIII}
\end{eqnarray}
Furthermore, in Sec.IV  the density matrix can be written, in the vector basis
$\{|G\rangle, |E_{\not k, \not l}\rangle, |k+l\rangle, |k,l\rangle, |B\rangle, |H\rangle\}$, as
\begin{eqnarray}
\rho=\left(\begin{array}{cccccc}
b_0 & 0 & 0 & 0 & 0 & 0 \\
0 & b_1 & b_3 & 0 & 0 & 0 \\
0 & b_3 & b_2 & 0 & 0 & 0 \\
0 & 0 & 0 & b_4 & b_7 & b_8 \\
0 & 0 & 0 & b_7 & b_5 & b_9 \\
0 & 0 & 0 & b_8 & b_9 & b_6
\end{array}\right).
\label{rhoIV}
\end{eqnarray}
Then, given the diagonal block structure of the above matrices \eqref{rhoIII} and \eqref{rhoIV}, we can argue that
for a number of initially excitations $e$ the density matrix  will have $e+1$ blocks of dimensions 
$1,2,\cdots, e+1$. Each block of dimension $r$ (with $1\le r \le e+1$) has
$\frac{r(r+1)}{2}$ number of real parameters. Since the total number of parameters in the matrix
determines the dimension of $\mathbb{H}_{\rho(0)}$, we will have
$dim \mathbb{H}_{\rho(0)}=\sum_{r=1}^{e+1}\frac{r(r+1)}{2}=(1+e)(2+e)(3+e)/6$ which is polynomial of degree 3 of $e$.

%%%%%%%%%%%%%%%%%%%%%%%%%%%%%%%%%%%%%%%%%%%%%%%%%%%%%%%%%%%%%

\appendix

\section{Solution of Eqs. (\ref{eqb0}) and (\ref{eqv})}

\begin{widetext}
\ba
b_0&=&\frac{8}{n-2}\left(\frac{e^{-4(n-1)t}-1}{4(n-1)}-\frac{e^{-2nt}-1}{2n}\right) ,\notag\\
b_1&=&\frac{4}{(n-2)n^2}(1-e^{-nt})^2(1-e^{-2(n-2)t}) ,\notag\\
b_2&=&-\frac{(1-e^{-2(n-2)t})(2e^{-nt}+n-2)^2}{(n-2)n^2},\notag\\
b_3&=&\frac{2(1-e^{-nt})(1-e^{-2(n-2)t})(n-2(1-e^{-nt}))}{(n-2)n^2} ,\notag\\
b_4&=&\left(\frac{2(n-1)e^{-(n-2)t}+2e^{-2(n-1)t}+n(n-3)}{n(n-1)}\right)^2,\notag\\
b_5&=&\frac{4e^{-4(n-1)t}}{(n-1)^2n^2}+\frac{4(n-4)e^{-(3n-4)t}}{n^2(n-1)(n-2)}
+\frac{(n-4)^2e^{-2(n-2)t}}{n^2(n-2)^2}-\frac{4(n-3)e^{-2(n-1)t}}{n(n-1)^2(n-2)}\notag\\
&&-\frac{2(n-3)(n-4)e^{-(n-2)t}}{n(n-1)(n-2)^2}+\left(\frac{(n-3)}{(n-1)(n-2)}\right)^2,\notag\\
b_6&=&4\left(\frac{e^{-2(n-1)t}}{n(n-1)}-2\frac{e^{-(n-2)t}}{n(n-2)}+\frac{1}{(n-1)(n-2)}\right)^2,\notag\\
b_7&=&4\frac{e^{-4(n-1)t}}{n^2(n-1)^2}+2\frac{(3n-8)e^{-(3n-4)}}{n^2(n-1)(n-2)}
+2\frac{(n-3)^2e^{-2(n-1)t}}{n(n-1)^2(n-2)}+2\frac{(n-4)e^{-2(n-2)t}}{n^2(n-2)}\notag\\
&&+\frac{(n-3)(n-6)e^{-(n-2)t}}{n(n-1)(n-2)}-\frac{(n-3)^2}{(n-1)^2(n-2)} ,\notag\\
b_8&=&4\left(\frac{e^{-4(n-1)t}}{n^2(n-1)^2}+\frac{(n-4)e^{-(3n-4)t}}{n^2(n-1)(n-2)}\right.\notag\\
&&\left.+\frac{(n^2-5n+8)e^{-2(n-1)t}}{2n(n-1)^2(n-2)}-2\frac{e^{-2(n-2)t}}{n^2(n-2)}
-\frac{(n-4)e^{-(n-2)t}}{n(n-1)(n-2)}+\frac{(n-3)}{2(n-1)^2(n-2)}\right) ,\notag\\
b_9&=&-2\left(-2\frac{e^{-4(n-1)t}}{n^2(n-1)^2}-\frac{(n-8)e^{-(3n-4)t}}{n^2(n-1)(n-2)}\right.\notag\\
&&\left.+\frac{(n-5)e^{-2(n-1)t}}{n(n-1)^2(n-2)}+2\frac{(n-4)e^{-2(n-2)t}}{n^2(n-2)^2}
-\frac{(3n-10)e^{-(n-2)t}}{n(n-1)(n-2)^2}+\frac{(n-3)}{(n-1)^2(n-2)^2}\right) .
\ea
\end{widetext}

%%%%%%%%%%%%%%%%%%%%%%%%%%%%%%%%%%%%%%%%%%%%%%%%%%%%%%%%%%%%%%%%%%%%%%%%%%%%%%%%%%%%%%%%%%%%%%%%%%%%%%%%%%%%%%%%%%%%%%%%%%%%%%%%%%%%%%%%%%%%%%%%%%%%%%%%%%%%%%%%%%%%%%%%%%%

\end{document}